\documentstyle[12pt]{article}

\def\sdspace{\baselineskip = .25in}
\def\dspace{\baselineskip = .30in}

\begin{document}

\begin{titlepage}
\begin{flushright} BA-94-45
\end{flushright}
\vspace{0.2cm}
\begin{center}{\LARGE Supersymmetric SO(10) Simplified\footnote{Supported
in part by Department of Energy Grant \#DE-FG02-91ER406267}}\\
\vspace{30pt}
{\large \bf K.S.Babu and S.M.Barr}\\ \vspace{0.1cm}
{\it Bartol Research Institute\\ University of Delaware\\
Newark, DE 19716}\vspace{0.3in}
\end{center}
\begin{abstract}
\sdspace

In the context of supersymmetric $SO(10)$ grand unified models,
it is shown that the gauge symmetry breaking as well as a natural
doublet--triplet splitting can be achieved with a minimal Higgs
system consisting of a single adjoint and a pair of vector and
spinor multiplets.  Such a Higgs spectrum has been shown to arise in
the free fermionic formulation of superstrings.  Since the
symmetry breaking mechanism relies on non--renormalizable
operators, some of the Higgs particles of the model turn out
to have masses somewhat below the GUT scale.  As a consequence, the
unification scale is raised to about $2 \times 10^{17}~GeV$ and
sin$^2\theta_W$ is predicted to be slightly larger than the minimal
SUSY--$SU(5)$ value.  Including threshold uncertainties, which turn
out to be surprisingly small in the model, we show that sin$^2\theta_W$
prediction is consistent with experiments.

\end{abstract}
\end{titlepage}
\newpage

\dspace

\section*{I. Introduction}
There has been a rebirth of interest in supersymmetric grand
unification since the improved measurements of the low energy
gauge couplings has confirmed that supersymmetry leads to an
astonishingly accurate unification of couplings [1].
The minimal SUSY--GUT prediction for $\sin^2 \theta_W$ is
$0.2334 \pm 0.0036$ to be compared with the experimental
value of $0.2324 \pm 0.0003$ [2]. (We have combined the experimental
and theoretical uncertainties in quadrature.)
It has long been believed by many theorists that
low energy supersymmetry is a necessary ingredient of grand
unification anyway if the gauge hierarchy problem is to be
solved in a satisfactory manner.

One aspect of the gauge hierarchy problem is the issue of
``doublet triplet splitting", ie., keeping
the pair of Higgs doublets of the supersymmetric standard
model light while giving their color triplet partners superheavy
masses to avoid excessive Higgs or Higgsino-mediated proton decay.
There are several mechanisms that have been proposed for achieving
doublet-triplet splitting in SUSY-GUTs without fine-tuning of
parameters: the ``sliding singlet mechanism" [3], the ``missing
partner mechanism" [4], the ``Dimopoulos-Wilczek mechanism" [5]
and the ``pseudo-goldstone-boson mechanism'' [6].
There is also the option of fine-tuning parameters of the
superpotenital to achieve doublet-triplet splitting.  The
non--renormalization theorem of supersymmetry will make such a
fine--tuning stable under radiative corrections.  Apart from
being aesthetically unappealing, it seems unlikely that the small
numbers required for this procedure $(\sim 10^{-14}$) would arise in a
more fundamental theory such as the superstring theory.

In SUSY $SU(5)$ the only way to do doublet triplet splitting
naturally is the missing partner mechanism [4], which requires
the existence of Higgs multiplets in the representations
${\bf 50}$, $\overline{\bf 50}$, and ${\bf 75}$. Aside from the lack
of economy involved in the introduction of these rank-four tensors,
it is questionable whether such high rank representations would be allowed
if the SUSY-GUT arises from an underlying superstring theory.

$SO(10)$ is in a number of ways a more attractive group for grand
unification. All the particles of a family are unified into a
single irreducible representation, the right--handed neutrino
automatically emerges, violation of $(B-L)$ allows the generation
of a sphaleron--proof cosmological baryon asymmetry, and anomaly
cancellation is automatic in $SO(10)$, among other things [7].

In $SO(10)$ it appears that the only possibility for natural
doublet--triplet splitting is the Dimopoulos-Wilczek mechanism [5].
This mechanism is quite simple. In its simplest form the masses of
the colored
Higgs(ino) fields arise from a term $T_1^a A^{ab} T_2^b$
(where the $T_i$ are ${\bf 10}$'s of $SO(10)$) with the
vacuum expectation value of the adjoint Higgs, $A^{ab}$, being
in the direction $\langle A \rangle =$ diag$(0,0,a,a,a) \otimes
(i \tau_2) \;\; \propto (B-L)$. This form of the vacuum expectation
value,
which we call the Dimopoulos-Wilczek (DW) form, gives a Dirac
mass assumed to be of order the GUT scale to the color ${\bf 3}
+ \overline{\bf 3}$ Higgsinos and Higgs, $\overline{\bf 3_1}
{\bf 3_2}$ and $\overline{\bf 3_2} {\bf 3_1}$, while leaving
the associated two pairs of Higgs(ino) doublets, $(\overline{\bf 2_1}
+ {\bf 2_1} + \overline{\bf 2_2} + {\bf 2_2})$ light.
With the additional term $M_2 (T_2)^2$ one of the pairs of
doublets, $\overline{\bf 2_2} + {\bf 2_2}$, can be made superheavy,
thus leaving the correct spectrum for the MSSM and preserving
the correct prediction of $\sin^2 \theta_W$. As was emphasized
in Refs. [8,9], Higgsino-mediated proton decay, which is a general
problem for SUSY GUTs, can readily be suppressed in this scheme
to acceptable levels by making $M_2$ only slightly
smaller than the VEV of $A^{ab}$.

In previous papers [8,9] we showed that it was possible to construct
a realistic and natural supersymmetric $SO(10)$ model with no fine-tuning
using the Dimopoulos-Wilczek mechanism. It was found
necessary in these papers to introduce the following
Higgs representations to do the breaking of $SO(10)$: ${\bf 54}$,
three ${\bf 45}$'s, ${\bf 16}$, and $\overline{\bf 16}$. The
${\bf 54}$ was required to give the DW form to the VEV of one
of the adjoints. The pair of spinors was needed to break $SO(10)$
all the way to the standard model gauge group. And the presence
of three adjoints was required to link the
two sectors (the ${\bf 54 + 45}$ to the $\overline{\bf 16} +
{\bf 16}$) in such a way as to avoid goldstone bosons and preserve
the DW form of the adjoint VEV. This will be explained further in
Section 2.

An important question is whether the Higgs spectrum needed for the
symmetry breaking as well as for
achieving a natural doublet--triplet splitting can
arise from an underlying superstring theory.  It has been known for some
time that conventional GUTs such as $SO(10)$ with scalars in the adjoint
representation can indeed arise in the free fermionic formulation of
superstrings [10].  Such a string construction requires the Kac--Moody level
to be two or higher.  In a recent paper, Chaudhury, Chung and Lykken [11]
have given an explicit level two string construction of SUSY
$SO(10)$ which has a single adjoint (along with arbitrary number of
${\bf \overline{16}}, {\bf 16}$ and {\bf 10}) that survive below the Planck
scale.  Furthermore, these authors have classified the
allowed representations that can emerge as massless chiral
multiplets below the Planck scale at the level two construction: the
number of adjoints is 0, 1 or 2, while the number of ${\bf 54}$ is
0 or 1.  No examples with more than one adjoint and/or one {\bf 54}
have so far been constructed.

The Higgs spectrum used for a natural doublet--triplet splitting in
Ref. [8,9], as it employs three adjoints, is not compatible with the
superstring construction of Ref. [11].  In this paper, we therefore wish
to address whether a realistic and natural
SUSY $SO(10)$ model can be constructed with only a single adjoint
Higgs.  In particular, we wish to do away with the ${\bf 54}$, so that
the spectrum will be identical to the explicit superstring construction
of Ref. [11].  From a group theory point of view, a single adjoint and a
${\bf \overline{16}} + {\bf 16}$ is sufficient to do the gauge symmetry
breaking.  We will show that it would also suffice to achieve a natural
doublet--triplet splitting via the DW mechanism.
In fact the Higgs content of the
model we shall construct is minimal: there are two ${\bf 10}$'s
and a ${\bf 45}$ as required for the Dimopoulos-Wilczek mechanism,
a $\overline{\bf 16} + {\bf 16}$ pair to complete the $SO(10)$
breaking, and a few singlets. As will be seen there is some
flexibility in the details of the model. However, certain features
are generic. In particular since the symmetry breaking mechanism relies
on non--renormalizable operators,
it turns out that there must be
certain multiplets of colored Higgs(inos) which are very light
compared to the GUT scale. As a result,
the unification scale $M_U$ is pushed above the minimal
SUSY-$SU(5)$ value to
about $2\times 10^{17}~GeV$.
This may be a welcome feature since $M_U$ is now closer
to the string compactification scale.  These scalars also
have an effect on $\sin^2\theta_W$ which is predicted in the model
to be somewhat larger
than the SUSY-$SU(5)$ value.  Though threshold
corrections turn out to be surprisingly small, they can be large
enough to somewhat compensate for the effects of these light
colored fields, thus making sin$^2\theta_W$ consistent with experiments.
Nevertheless, one expects that $\sin^2 \theta_W$
will be somewhat on the large side in SUSY $SO(10)$ if the particle
content is as currently suggested by superstring theory.

\section*{II. The Model}

\noindent
{\bf A. Achieving the DW form without a 54.}

In Refs. 8 and 9 the required Dimopoulos-Wilczek form
of the adjoint vacuum expectation value was achieved by
means of versions of the following superpotential, which
we called the Srednicki sector [12].
\begin{equation}
W = m_1 A^2 + \lambda_1 S A^2 + m_2 S^2 + \lambda_2 S^3.
\end{equation}

\noindent
Here $A$ is an adjoint and $S$ is a ${\bf 54}$.
If it is assumed that $\langle A \rangle$ is of the form
diag$(b,b,a,a,a) \otimes (i \tau_2)$, then the $F_A =0$
equation gives two conditions,
\begin{equation}
\begin{array}{ccl}
(m_1 - \frac{3}{2} s) b $ = $ 0,\\
(m_1 + s) a $ = $ 0,
\end{array}
\end{equation}

\noindent
where the VEV of $S$ is $\langle S \rangle =$ diag$(-\frac{3}{2} s,
-\frac{3}{2} s, s,s,s) \otimes I$. Clearly, one solution of this
is $b = 0$ and $a \neq 0$ which is the DW form with $s = -m_1$.

If there is no ${\bf 54}$ in the model, and the only tensors
are one or more adjoints, then it appears that there is only
one way to achieve the DW form, and that involves
higher--dimension operators. In particular, with only a single adjoint
one may write down
\begin{equation}
W = m (tr A^2) + \beta \frac{1}{M_{Pl}}(tr A^2)^2 + \beta^{\prime}
\frac {1}{M_{Pl}} (tr A^4).
\end{equation}

\noindent
This is easily seen to have as a solution $b = 0$ and
$a = \sqrt{m M_{Pl}/ (12 \beta + 2 \beta^{\prime})}$.
Notice that if the VEV of $A$ is to be of order
the GUT scale, $M_G$, then the mass parameter of $A$ has to
be of order $M_G^2/M_{Pl}$.
To explain the appearance of the higher-dimension operators
it shall be assumed here and in what follows that all
operators not forbidden by local symmetry will be induced
by Planck-scale physics suppressed only by the dimensionally
appropriate powers of the Planck mass, $M_{Pl} = 1.2 \times
10^{19}$ GeV.\footnote{The proper expansion parameter in superstring
theories may turn out to be the ``reduced Planck mass''
$\simeq  2 \times 10^{18}~GeV$ rather than $M_{Pl}$.
This variance can be readily accommodated
into our analysis by correspondingly reducing the
coefficients of the non--renormalizable operators (which are treated as
free parameters).}
 This is what is generally expected in string theory.

\noindent
{\bf B. Breaking $SO(10)$ to the Standard Model Group.}

The VEV of the adjoint Higgs having the DW form breaks
$SO(10)$ down to $SU(3)_c \times U(1) \times SO(4)$. To
complete the breaking to the Standard Model group requires
a $\overline{\bf 16} + {\bf 16}$, which will be denoted by
$\overline{C} + C$. What is required is that the components
of $C$ and $\overline{C}$ that are singlets under the
Standard Model group acquire VEVs of order $M_G$. This is
achievable in a number of ways. For example, there could be a singlet
superfield, $X$, with coupling $X(\overline{C} C - M^2)$.
Another possibility, similar to the adjoint sector, is
to choose

\begin{equation}
W = m \overline{C} C + \beta \frac {1}{M_{Pl}} (\overline{C} C)^2
+ \beta^{\prime} \frac{1}{M_{Pl}} (\overline{C} \gamma ^{ab} C)^2.
\end{equation}

\noindent
This is the most general form the superpotential for
$C$ and $\overline{C}$ can take up to fourth order.  (Terms such as
$C^4$ and $\overline{C}^4$ do not play a role in symmetry breaking.)

\noindent
{\bf C. The problem of linking the adjoint and spinor sectors.}

There are, as just seen, two sectors of Higgs fields necessary
to do the
breaking of $SO(10)$, one containing the adjoint which gets
a VEV in $(B-L)$ direction (the Dimopoulos-Wilczek form),
and a sector containing the pair of spinor fields which is
needed to break the rank of the group to four and complete
the breaking to the Standard Model group. A crucial problem
in $SO(10)$ is how to link these sectors. They must certainly
be linked, for otherwise there is nothing in the superpotential
to determine the relative orientation of the VEVs in the two
sectors. The result would be the existence of very light
pseudogoldstone particles corresponding to those generators
that are broken in both sectors. These are easily found to be
in the representations $[{\bf (3,2,\frac {1}{6})} + {\bf (3,1,
\frac {2}{3})} +$ h.c.] of the Standard Model group.

The reason that this is a problem is that terms that couple the
two sectors will, in most cases, destabilize the Dimopoulos-Wilczek
form of the adjoint VEV. For example, a $\overline{C} \gamma^{ab}
C A^{ab}$ term leads to a linear term in $b$ where
$\langle A \rangle =$ diag$(b,b,a,a,a) \otimes (i \tau_2)$.

The way this problem was solved in Refs. 8 and 9 was through
a term of the form tr$(A A^{\prime} A^{\prime \prime})$. Here
$A$ is the adjoint which has a VEV in the DW form, and $A^{\prime
\prime}$ is an adjoint which couples to $\overline{C} C$ and
whose VEV thus does not have the DW form. $A^{\prime}$ is a
third adjoint which is necessary because of the fact that
tr$(A A^{\prime} A^{\prime \prime})$ is a totally antisymmetric
term and therefore will vanish if any two of the adjoints are
the same.  Because of its
total antisymmetry this term does not affect the VEVs of the fields.
[This is easily seen from its contribution to $F_A^{ab}$, which
is $(\langle A^{\prime} \rangle \langle A^{\prime \prime} \rangle)
^{ba} = 0$, because the indices $[a,b]$ are antisymmetrized.
It is assumed here that the VEVs of the adjoints are all
in normal form: $(a_1,a_2,a_3,a_4,a_5) \otimes (i \tau_2)$.]
However, this term does contribute to the masses of the
would-be goldstone bosons. The drawback of this elegant term
is that it of necessity involves three distinct adjoints,
while the present goal is to find a way to make do with
only one or at most two adjoints [11].

It might be supposed that the same trick would work with
the role of the $A^{\prime}$ and $A^{\prime \prime}$ being
played by spinor-antispinor pairs contracted to form adjoints:
$(\overline{C} \gamma^{ab} C) (\overline{C'} \gamma^{bc} C')
A^{ca}/M_{Pl}^2$. While this term would indeed not destabilize
the DW form of $\langle A \rangle$, that is not the case
with other terms which can be obtained by contracting the same
fields in different ways as, for example, $(\overline{C}
\gamma^{ab} C) (\overline{C}^\prime C^\prime) A^{ab}/M_{Pl}^2$. It is
easily seen that there is no abelian symmetry which can allow
one contraction of the fields while ruling out others.
(With non-abelian discrete symmetries this can be done, but it is
not clear that the necessary symmetries can emerge from string
theory, and the examples we have found seem quite contrived,
so we will not present them.)

In sum, there seems to be no way to link the two sectors
without destabilizing the DW form, unless there are either three
adjoints or non-abelian discrete symmetries that are respected
by higher-dimension operators.

It seems that the only acceptable possibility is that the DW form
is, indeed, destabilized, but only by a small enough amount
that the gauge hierarchy is not destroyed. How small is small
enough can be determined by examining the two-by-two
matrices for the Higgs(ino) masses. These come from
terms of the form $\lambda T_1^a A^{ab} T_2^b + M_2 (T_2)^2$.
If the Higgs that couple to light quarks and leptons are in
$T_1$, then the Higgsino-mediated proton decay amplitude is
proportional to $({\cal M}^{-1})_{11} =  M_2/(\lambda a)^2$
where ${\cal M}^{-1}$
is the inverse of the two-by-two mass matrix of the colored
Higgsinos. On the other hand, the $\mu$-parameter of the
light Higgs(ino) doublets in $T_1$ receive a contribution
$\delta \mu = (\lambda b)^2/M_2$. Thus
\begin{equation}
\delta \mu \cdot ({\cal M}^{-1})_{11} = (b/a)^2.
\end{equation}

\noindent
A comfortable agreement with the experimental limits on
proton decay requires that $({\cal M}^{-1})_{11} \stackrel{_<}
{_\sim} (10^{17}$GeV$)^{-1}$, while naturalness of the gauge
hierarchy requires that $\delta \mu  \stackrel{_<}{_\sim}
1$ TeV. Thus, one requires that
\begin{equation}
b/a \stackrel{_<}{_\sim} 10^{-7}.
\end{equation}

\noindent
Such a small VEV for $b$ can be achieved if the terms that
destabilize the DW form by providing a linear term in $b$
are high order and thus suppressed by powers of $1/M_{Pl}$.
The price that is paid for this is that the masses of the
pseudogoldstone bosons that arise from the same
higher--dimension terms will be also very small compared to the
GUT scale, as will be seen. This will be reflected in
$\sin^2 \theta_W$ as well as in the unification scale.

\noindent
{\bf D. A model of the $SO(10)$--breaking sector.}

In order that lower order terms that would disrupt the
DW form and destroy the gauge hierarchy not be present
it is necessary that there be symmetries. In the
illustrative model of an $SO(10)$--breaking sector now to be
presented the symmetry is
a $Z_4 \times Z_4$. The relevant fields are an adjoint, $A$,
a spinor--antispinor pair, $\overline{C} + C$, and two pairs of
singlets, $(P_1$, $\overline{P_1})$, and $(P_2, \overline{P_2})$.
Under the first $Z_4$, $A \rightarrow
i A$, $P_1 \rightarrow i  P_1$, $\overline{P_1} \rightarrow -i
\overline{P_1}$, and the other fields transform
trivially. Under the second $Z_4$, which will be denoted $Z_4^{\prime}$,
$\overline{C} \rightarrow i \overline{C}$, $C \rightarrow i C$,
$P_2 \rightarrow i P_2$, $\overline{P_2} \rightarrow -i \overline{P_2}$,
and the other fields transform trivially.

Under these symmetries the most general superpotential (up to
the relevant orders in $M_{Pl}^{-1}$) is given by
\begin{eqnarray}
W & = & \frac{1}{M_{Pl}}(\alpha_1 P_1^2 + \overline{\alpha_1}
\overline{P_1}^2)
tr(A^2)  +  \frac{\beta_1}{M_{Pl}}(tr(A^2))^2
 +  \frac{\beta_1^{\prime}}{M_{Pl}} tr(A^4) \\ \nonumber
& & + \; m_1 P_1 \overline{P_1}  +  \frac{1}{M_{Pl}}(\gamma_1 P_1^4
+ \overline{\gamma_1} \overline{P_1}^4 + \gamma_1^{\prime} P_1^2
\overline{P_1}^2) \\ \nonumber
& & + \frac{1}{M_{Pl}}(\alpha_2 P_2^2 + \overline{\alpha_2}
\overline{P_2}^2) \overline{C} C  +  \frac{\beta_2}{M_{Pl}}
(\overline{C} C)^2  +  \frac{\beta_2^{\prime}}{M_{Pl}}
(\overline{C} \gamma^{ab} C)^2 \\ \nonumber
& & + \;  m_2 P_2 \overline{P_2}  +  \frac{1}{M_{Pl}}
(\gamma_2 P_2^4 + \overline{\gamma_2}\overline{P_2}^4+\gamma_2^\prime
P_2^2\overline{P_2}^2) \\
& & +  \sum_i \frac{\delta_i}{M_{Pl}^5} \{ (\overline{C} C)^2 A^4  +
(\overline{C} C P_2^2) A^4  +  (\overline{C} C)^2 (A^2 P_1^2)  +
(\overline{C} C P_2^2) (A^2 P_1^2) \}_i. \nonumber
\end{eqnarray}

\noindent
Each term in the curly brackets on the right-hand side of Eq. (7) actually
corresponds to several terms contracted in different ways.
For example, there are seven distinct ways to contract
$(\overline{C} C)^2 A^2$: $(\overline{C} \gamma^a \overline{C})
(C \gamma^a C) tr(A^2)$, $(\overline{C} \gamma^a \overline{C})
(C \gamma^b C) A^{ac} A^{cb}$, $(\overline{C} \gamma^a \overline{C})
(C \gamma^{abcde} C) A^{bc} A^{de}$,
$(\overline{C} \gamma^{abcde} \overline{C})(C \gamma^a C) A^{bc} A^{de}$,
$(\overline{C} \gamma^{abcde} \overline{C})(C \gamma^{abcde} C) tr(A^2)$,
$(\overline{C} \gamma^{abcde} \overline{C})(C \gamma^{abcfg} C) (A^{de}
A^{fg}$ or $A^{df} A^{eg})$. Moreover, in each term in the curly
brackets $P_j^2$ can be replaced by $\overline{P_j}^2$.
The many terms in the curly brackets
are distinguished by the index `$i$', and each has a distinct
coefficient `$\delta_i$'.

All the dimensionless coefficients in
this superpotential, $\alpha_1$, $\beta_1$, ... , $\delta_i$, are
assumed to be of order unity. If the dimensionful parameters
$m_1$ and $m_2$ are assumed to be of order $M_G^2/M_{Pl}$, then
all the VEVs (except $b$) are of order $M_G$.

Defining $\langle \overline{C} \rangle = \langle C \rangle
\equiv c$, then $\langle P_2 \rangle \sim \langle \overline{P_2}
\rangle \sim c$, and $c$ sets the scale of $SO(10)$ breaking
to $SU(5)$. Recalling that $\langle A \rangle \equiv$ diag$(b,b,a,a,a)
\otimes (i \tau_2)$, then $\langle P_1 \rangle \sim \langle
\overline{P_1} \rangle \sim a$, and $a$ sets the scale of $SU(5)$
breaking to the Standard Model group.

The VEV $b$ is determined by the $F_A = 0$ equation, which gives
\begin{equation}
b/a \approx (\delta_{eff}/\beta_1^{\prime}) \left (\frac{c}{M_{Pl}}
\right)^4.
\end{equation}

\noindent
Here $\delta_{eff}$ is some linear combination of all the
$\delta_i$ that appear in the superpotential (Eq. (7)). If $b/a$ is
to be less than or of order $10^{-7}$ then $c/M_{Pl}$ must
be less than or of order $2 \times 10^{-2}$, which, as shall
be seen later from solving the renormalization group equations
for $M_G$, is reasonable.

This shows the importance of suppressing by discrete symmetry
lower dimension operators
linking the adjoint and spinor Higgs fields. If, for example,
a term $\overline{C} C A$ were allowed,
it would need to have a coefficient of order $10^{-9}$
to make $b/a \stackrel{_<}{_\sim} 10^{-7}$, which is a fine-tuning.
By having the first operator which destabilizes the DW form
of $\langle A \rangle$ be  $O(1/M_{Pl}^5)$, it is possible to have
a realistic model with all dimensionless parameters being of order
unity.

\dspace

\noindent
{\bf E. The doublet-triplet splitting.}

For reasons that will become apparent below, assume that the
Higgs(ino) masses come from higher dimension operators.

\begin{equation}
W(T_1,T_2) = \lambda T_1^a A^{ab} T_2^b \left(\frac{P}{M_{Pl}}\right)^n
+ \lambda^{\prime}(T_2)^2\left(\frac{Q^{2n+1}}{M_{Pl}^{2n}}
\right).
\end{equation}

\noindent
Then the doublet and triplet Higgsino mass matrices are given
by

\begin{equation}
T_1 M T_2 = ({\bf \overline{2_1}},{\bf \overline{2_2}})
\left( \begin{array}{cc}
0 & i \Lambda b \\ -i \Lambda b & M_2 \end{array} \right)
\left( \begin{array}{c} {\bf 2_1} \\ {\bf 2_2} \end{array} \right) +
({\bf \overline{3_1}}, {\bf \overline{3_2}}) \left( \begin{array}{cc}
0 & i \Lambda a \\ -i \Lambda a & M_2 \end{array} \right)
\left( \begin{array}{c} {\bf 3_1} \\ {\bf 3_2} \end{array} \right),
\end{equation}

\noindent
where $M_2 = \lambda^{\prime}(\langle Q \rangle ^{2n+1}/M_{Pl}^{2n})$
and $\Lambda = \lambda(\langle P \rangle /M_{Pl})^n$. Thus the contribution
to the $\mu$ parameter is
\begin{eqnarray}
\delta \mu & = & \Lambda^2 b^2/M_2 = \left(\frac{\lambda^2}{\lambda
^{\prime}}\right)\frac{b^2 \langle P \rangle ^{2n}}{\langle
Q \rangle ^{2n+1}} \nonumber \\
& \sim & b^2/M_G \sim 10^{-14} a^2/M_G \sim 10^{-14} M_G \nonumber \\
& \sim & 10^3 ~GeV,
\end{eqnarray}

\noindent
where it has been assumed that the VEVs of $P$ and $Q$
are, like all the other VEVs, of order $M_G$, and the dimensionless
couplings $\lambda$ and $\lambda^{\prime}$, like all the others,
are of order unity.

The parameter that controls Higgsino-mediated proton decay,
$({\cal M} ^{-1})_{11}$ is given by
\begin{equation}
({\cal M}^{-1})_{11} = M_2/(\Lambda ^2 a^2) = (\frac{\lambda^{\prime}}
{\lambda^2}) \frac {\langle Q \rangle^{2n+1}}
{a^2 \langle P \rangle ^{2n}} \sim M_G^{-1}.
\end{equation}

\noindent
This is the correct order if proton decay is to be suppressed
to realistic levels. Note that Eqs. (5) and (6) are satisfied.

The integer $n$ appearing in the powers of $M_{Pl}$ in Eq. (9) is determined
by the following consideration. Any symmetry that allows the terms
in Eq. (9) will also allow a term of the form $(T_1)^2 A^2 P^{2n}
({\bar Q})^{2n+1}/M_{Pl}^{4n+2}$, assuming there is a chiral
superfield $\overline{Q}$ with the opposite quantum numbers
to $Q$.  Since it has been assumed throughout that Planck-scale physics
induces, unsuppressed except by powers of $M_{Pl}$,
all higher-dimension operators allowed by local symmetry, it must
be assumed that this operator exists also in the effective
sub-Planck-scale theory. This would give a contribution to the
$\mu$ parameter of order $M_G(M_G/M_{Pl})^{4n+2}$. It will be
seen later that $M_G/M_{Pl} \approx \frac{1}{45}$, so that to
avoid destroying the gauge hierarchy $n$ must be $\geq 2$.
One can impose a local U(1) symmetry that guarantees the form
of Eq. (9) with no lower dimension operators contribute to
$\mu$.  For example, if $n=2$, the $U(1)$ charges of $(P,Q,T_1,T_2)$
can be chosen to be $(2,-2,5,9)$ with all the remaining fields having
zero charge.  Instead of a $U(1)$ symmetry it is
possible to use a discrete subgroup of the $U(1)$ (eg: $Z_9$
or $Z_{18}$ in the
above example).  As shown in Ref. 9, it is straightforward to make these
symmetries free of anomalies so that they are ``local''.

\noindent
{\bf F. The spectrum of the model.}

Of the 45 gauge bosons of $SO(10)$, 12 are the gauge bosons
of the Standard Model and remain light. The rest have masses
of order $M_G$. In particular, $M^2({\bf (3,1,\frac{2}{3})} + h.c.)
= 4 g^2 (c^2 + a^2)$, $M^2({\bf (3,2,\frac{1}{6})} + h.c.) =
g^2 (4 c^2 + a^2)$, $M^2({\bf (1,1, \pm 1)}) = M^2({\bf (1,1,0)}) = 4 g^2 c^2$,
and $M^2({\bf (3,2,-\frac{5}{6})} + h.c.) = g^2 a^2$.

Of the 77 $(= {\bf 45} + {\bf 16} + {\bf \overline{16}})$ Higgs(ino) components
involved in the breaking of $SO(10)$ to the Standard Model, 33 are
eaten to the give the massive gauge multiplets just enumerated.
11 components of the ${\bf 16} + {\bf \overline{16}}$ (namely ${\bf
\overline{5}} +
{\bf 5 }+ {\bf 1}$ under $SU(5)$) acquire mass of order $c^2/M_{Pl} \sim m_2$.
These fields have masses that are very nearly $SU(5)$ invariant
because their coupling to $\langle A \rangle$ is so weak
(ie.$O(M_G^6/M_{Pl}^5)$). They therefore have a negligible effect
on $\sin^2 \theta_W$.

Further, 15 components of the adjoint, $A$, acquire masses of order
$a^2/M_{Pl} \sim m_1$. In particular, $M({\bf (1,1,0)}) = M({\bf (8,1,0))
}= 4 \beta_1^{\prime} a^2/M_{Pl}$, $M({\bf (1,3,0)}) = M({\bf (1,1, \pm 1)})
= M({\bf (1,1,0)}) = 2 \beta_1^{\prime} a^2/M_{Pl}$.

Finally, there are 18 pseudogoldstone bosons (and their fermionic
partners) that come from both $A$ and $\overline{C} + C$. Their
masses are $M({\bf (3,1, \frac{2}{3})} + h.c.) = \delta_{eff}^{(1)}
a^2 c^2 (c^2 + a^2)/M_{Pl}^5$, and $M({\bf (3,2,\frac{1}{6})} + h.c.)
= \delta_{eff}^{(2)} a^2 c^2 (4c^2 + a^2)/M_{Pl}^5$, where
$\delta_{eff}^{(1)}$ and $\delta_{eff}^{(2)}$ are some linear
combinations of the $\delta_i$ appearing in the superpotential.
With $M_G/M_{Pl} \sim \frac{1}{45}$ (as shall be found later)
these pseudogoldstones have masses of order $2 \times 10^9$ GeV.

In the ${\bf 10}$'s, $T_1$ and $T_2$, there are two $[{\bf (3,1,-\frac{1}{3})
}+ h.c.]$ pairs, the product of whose masses is seen from Eq. (10)
to be $\Lambda^2 a^2 \sim (M_G^3/M_{Pl}^2)^2$. There is one
pair of ${\bf (1,2,-\frac{1}{2})} + h.c.$ with mass $M_2$, and one
light pair with mass of order the weak scale.

\noindent
{\bf G. Realistic Fermion masses.}

The three families of quarks and leptons belonging to {\bf 16} of
$SO(10)$ (denoted by $F_I$, $I=$1-3) have the following
Yukawa couplings to $T_1$ and $\overline{C}$:
\begin{equation}
\lambda_{IJ} F_I F_J T_1 + \lambda_{IJ}^{\prime}
F_I F_J \left(\overline{C}.\overline{C}N/M^2_{Pl}\right)~,
\end{equation}
where $N$ is a gauge
singlet.  Clearly such terms respect all the symmetries of the model
discussed earlier.  The coupling to $T_1$ gives rise to the Dirac masses
of all fermions, while the coupling to $\overline{C}$ results in heavy
Majorana neutrino masses for $\nu_R$'s.  In order to correct the
bad $SU(5)$ mass relations, it is necessary for the light quark and
lepton masses to depend on the breaking of $SU(5)$.  Therefore they must
couple to $A$.  Of course, the direct coupling of $A$ to $F_IF_J$ is not
allowed by $SO(10)$.  One idea that has been suggested in the literature
[13,9] is that there are additional vector--like representations of quarks
and leptons.  If, for example, there is a ${\bf 16}+{\bf \overline{16}}$
(denoted by $F+\overline{F}$), then $A$ may couple as follows:
\begin{equation}
\lambda_I F_I \overline{F} \left({{AN}\over{M_{Pl}}}\right) +
mF \overline{F}~.
\end{equation}
This allows realistic quark and lepton mass relations [13,9].
Note that the gauge hierarchy is unaffected by these
vector fermions.

\newpage
\noindent
{\bf H. Cosmology of the pseudogoldstone bosons.}

As noted in II.F, the model has 18 pseudogoldstone bosons (and their
fermion partners) belonging to $[{\bf (3,1,{2\over 3})} + {\bf
(3,2,{1 \over 6})} +
h.c.]$ under the Standard Model gauge group.  They have masses of order
$2 \times 10^9~GeV$.  Since all the gauge bosons with which they
interact have masses of order
$M_U \sim 2 \times 10^{17}~GeV$, it is important to check if these
pseudogoldstones are so long--lived as to cause problems for cosmology.

The ${\bf (3,2,{1\over6})}$ (denoted by $\chi_1$)
pseudogoldstone can readily decay into
light fermions using the $T_1^aA^{ab}T_2^b$ interaction of Eq. (9).
$\chi_1$ decays into a light doublet from $T_1$ and a heavy (virtual)
color triplet from $T_2$.  By using the same $T_1AT_2$ vertex, the
color triplet in $T_2$ converts into a color triplet in $T_1$, which
has Yukawa couplings to the light quarks and leptons.  The amplitude for
this decay $\chi_1 \rightarrow H_1F_IF_J$
goes as $1/M_G$ with the decay rate $\Gamma_d \sim m_{\chi_1}^3
/M_G^2\sim 10^{-6}~GeV$.  Comparing $\Gamma_d$ with the expansion
rate of the universe $\Gamma_{exp} \sim T^2/M_{Pl}$, we see that the
freeze--out temperature is $T_* \sim 10^7~GeV$, which is
sufficiently high and quite safe.

As for the ${\bf (3,1,2/3)}$ pseudogoldstone (denoted by $\chi_2$),
there is no direct coupling with
the light Higgs in $T_1$.  However, it also decays quite fast.  The
interaction of $A$ listed in Eq. (14), along with the superpotential
terms $\alpha A^2 P^2$ and $\beta A^4$ (see Eq. (7)) lead to the
$D$--term
\begin{equation}
\lambda_I \lambda_J
\overline{F_I}\partial_\mu \gamma^{\mu} F_J A \left({1 \over {16 \pi^2}}
{{\langle A \rangle\alpha \beta}\over {M_G^2}}\right)~,
\end{equation}
which arises through a one-loop diagram.  The decay rate for
$\chi_2 \rightarrow F_I F_J$ is then $\Gamma_d \sim
\left({1 \over {16 \pi^2}} {{\lambda_I \lambda_J\alpha \beta}
\over {M_G}}\right)^2
m_{\chi_2}^3 \sim (\lambda_I\lambda_J\alpha \beta )^2
\times 10^{-11}~GeV$.  The
corresponding freeze--out Temperature is $T_* \sim
(\lambda_I\lambda_J\alpha \beta) \times 10^4~GeV$, which is also quite safe.

It is easy to verify that these pseudogoldstone bosons do not mediate
proton decay at an unacceptable level.  The effective interaction of Eq.
(15) leads to a proton decay amplitude proportional to the light quark
masses (as in the usual dimension 5 proton decay of SUSY-GUT) and
a factor $1/(16 \pi^2 M_G)$.  This rate is negligibly small.  Similarly, box
diagrams with internal particle being $F,\overline{F}$ can be seen to
have an amplitude $\sim 1/M_{Pl}^2$ with the usual light quark Yukawa
suppression factors, which is also small.

\section*{III. The calculation of sin$^2\theta_W$}

For a number of reasons the uncertainties in $\sin^2 \theta_W$
due to physics at large scales are relatively quite small in
this model compared to what one might expect in $SO(10)$.
First, the string-theory-motivated constraints that have
been imposed have meant that there is only one large representation
of Higgs(inos), namely the adjoint. Second, because of the
extremely weak coupling of the spinor-antispinor pair of
Higgs(inos) to the adjoint, they contribute negligibly to
$\sin^2 \theta_W$ as already noted, since they are almost
perfectly $SU(5)$ degenerate. And, third, the gravitational
contributions are small since the only possible term of
order $M_{Pl}^{-1}$, namely tr$\frac{\langle A \rangle}{M_{Pl}}
F_{\mu\nu} F^{\mu\nu}$ vanishes
because of the antisymmetry of $A$.  The terms of order $M_{Pl}^{-2}$
will produce an uncertainty, at most, of order $10^{-3}$ in
$\sin^2 \theta_W$.

Moreover, the presence of the pseudogoldstones at intermediate
scales has the effect of somewhat pushing up $\sin^2 \theta_W$.
One expects, therefore, that $\sin^2 \theta_W$ will lie at the
high end of the presently allowed range.  This will be quantified
shortly.

The $SU(5)$ gauge bosons that mediate proton decay (the ${\bf (3,2,
-\frac{5}{6})} + h.c.$) have mass $ga$ which will be defined to be
$M_G$. The breaking of $SO(10)$ to $SU(5)$ contributes to the
mass-squared of the gauge bosons an amount $(2gc)^2$ which
will be denoted $M_{10}^2$. It will be convenient to define
$x \equiv M_{10}/M_G = 2c/a$.

{}From the one-loop renormalization group equations and using the spectrum
of particles listed in Section 2, we arrive at
\begin{eqnarray}
{\rm ln}({{M_U}\over {M_{Pl}}}) &=& {{\pi}\over {17 \alpha}} -{{8 \pi}\over
{51 \alpha_3}} - {1 \over 17} {\rm ln}(\rho_1^2\rho_5) + {10 \over 17}
{\rm ln}({{M_Z}\over {M_{Pl}}}) \nonumber \\
\alpha_G^{-1} &=& {7 \over 17} \alpha^{-1} -{5 \over 51} \alpha_3^{-1}
-{48 \over {17 \pi}}{\rm ln}\rho_1 + {37 \over {34\pi}}{\rm ln}\rho_5
+{89 \over {34\pi}}{\rm ln}({{M_Z}\over {M_{Pl}}}) \nonumber \\
{\rm sin}^2\theta_W(M_Z) &=& {5 \over 34} + {31 \over 51}{{\alpha}\over
{\alpha_3}} + {{5 \alpha}\over {34 \pi}}{\rm ln}({{\rho_1}\over
{\rho_5}}) + {{\alpha}\over {17 \pi}}{\rm ln}{\rho_1}
-{{9 \alpha}\over{17 \pi}}{\rm ln} ({{M_Z}\over {M_{Pl}}})
\end{eqnarray}

\noindent Here $\rho_1 = (2\beta_1^\prime/g^2)(M_G/M_U)^2$
is the coefficient of the Higgs(ino) mass from
the {\bf 45} which is of order $M_G^2/M_{Pl}$,
and $\rho_5 = \delta^{(1)}_{eff}{{x^2}\over 4}(1+{{x^2}\over 4})
({{M_G}\over{gM_U}})^6$ is the coefficient
of the Higgs(ino) pseudogoldstone multiplet which has a mass of order
$M_G^6/M_{Pl}^5$.  In the above, we have ignored the
mass--splitting between various multiplets that are of the same
order.  This will be treated as part of the threshold corrections.
$M_G = ga$ is the mass of the $SU(5)$ gauge boson and $M_U$ is the scale
at which the three couplings unify.

Using $\alpha_3(M_Z) = 0.12$ and $\alpha(M_Z) = 1/127.9$ as inputs,
we see that $\alpha_G^{-1} = 19$, or $g=0.81$ corresponding to
$\rho_1 = \rho_5 = 1$.  The unification scale is found to be
\begin{equation}
M_U/M_{Pl} \cong (2.3 \times 10^{-2}) [x^2(1+\frac{x^2}{4})]
^{-\frac{1}{17}}.
\end{equation}
For $x$ of order unity, $M_U/M_{Pl} \simeq 1/45$.
The one--loop prediction for sin$^2\theta_W$, ignoring threshold
effects for now, is
\begin{equation}
{\rm sin}^2\theta_W|_{1-loop} = 0.2384 - {{5 \alpha}\over{34\pi}}{\rm ln}
\left({{\delta^{(1)}_{eff}x^2} \over {8g^4\beta_1^\prime}}(1+{{x^2}\over 4})
\right) +
{{\alpha}\over {17 \pi}}{\rm ln}({{2\beta_1^\prime}\over {g^2}})
-{{8 \alpha}\over{17\pi}}{\rm ln}({{M_G}\over{M_U}})~.
\end{equation}

\noindent
The logarithmic terms reflect
the contribution of the light pseudogoldstones. One can estimate
the ratio $\delta_{eff}^{(1)}/\beta_1^{\prime}$ by considering
the ratio of VEVs $b/a$. From Eqs. (6) and (8) one has that
$\delta_{eff}/\beta_1^{\prime} \stackrel{_<}{_\sim}
10^{-7}\left(\frac{x}{2g}\frac{M_G}{M_{Pl}}\right)^{-4}$.
This gives $\ln(\delta_{eff}/\beta_1^{\prime}) \stackrel
{_<}{_\sim} - \ln(\frac{x^4}{4})$.  Thus the second term in Eq. (18)
tends to be a $positive$
contribution to sin$^2\theta_W$.  The $\delta_{eff}$
appearing in the expression for $b/a$ is not the same linear
combination of the $\delta_i$ that appears in the pseudogoldstone
masses and that has been denoted $\delta_{eff}^{(1)}$. However,
if all the $\delta_i$ are assumed to be comparable, the
difference as far as $\sin^2 \theta_W$ is concerned should
be negligible.

Let us now turn to the two--loop and threshold corrections to
sin$^2\theta_W$ in the model.
The two-loop correction to $\sin^2 \theta_W$  (including a conversion
factor to go from $\overline{MS}$ to $\overline{DR}$ scheme) is obtained
numerically to be
\begin{equation}
\sin^2 \theta_W|_{2-loop} = +0.0037.
\end{equation}

\noindent
The correction arising from the splitting among the superheavy
gauge multiplets is
\begin{equation}
\Delta \sin^2 \theta_W|_{gauge} = -\frac{\alpha}{10 \pi}[6 \ln x
+ 15 \ln (4 + x^2)^{\frac{1}{2}} -21 \ln (1 + x^2)^{\frac{1}{2}}].
\end{equation}

\noindent
The correction from the splittings among Higgs(ino) multiplets
with masses of order $a^2/M_{Pl}$ is
\begin{equation}
\Delta \sin^2 \theta_W|_{Higgs} = +\frac{\alpha}{30 \pi}
(21 \ln 2).
\end{equation}

\noindent
The correction coming from the splitting between the pseudogoldstone
multiplets is given by
\begin{equation}
\Delta \sin^2 \theta_W|_{pseudos} = \frac{\alpha}{30 \pi} (-21)
\left( \ln \left( \frac{1+x^2}{1+x^2/4} \right) + \ln \left(
\frac{\delta_{eff}^{(2)}}{\delta_{eff}^{(1)}} \right) \right).
\end{equation}

\noindent
The correction, finally, from the splittings among the
heavy fields in $T_1$ and $T_2$ is given by
\begin{equation}
\Delta \sin^2 \theta_W|_{T_i}  =  \frac{\alpha}{30 \pi}
9 \ln \left( \frac{\Lambda^2a^2}{M_2M_G} \right)
=  - \frac{\alpha}{30 \pi} 9 \ln (M_G ({\cal M}^{-1})_{11}).
\end{equation}

\noindent
Combining all the contributions one finds that
\begin{eqnarray}
{\rm sin}^2\theta_W(M_Z) &=& {5 \over 34} + {31 \over 51} {{\alpha}\over
{\alpha_3}}- {{9\alpha}\over{17\pi}}{\rm ln}({{M_Z}\over {M_{Pl}}})
- {{19 \alpha}\over {170\pi}}{\rm ln}2 + {{18 \alpha}\over{17\pi}}
{\rm ln}|g| + 0.0037 \nonumber \\
&+& {{\alpha}\over {17 \pi}}{\rm ln}(2\beta_1^\prime)
-{{5 \alpha}\over{34\pi}}{\rm ln}\left({{\delta^{(1)}_{eff}x^4}
\over {2\beta_1^\prime}}\right) - {{231 \alpha}\over {170 \pi}}{\rm ln}(
{{M_G}\over{M_U}}) \nonumber \\
&-&{{21 \alpha}\over{30\pi}}{\rm ln}\left({{\delta^{(1)}_{eff}}\over
{\delta^{(2)}_{eff}}}\right) - {{3\alpha}\over{10\pi}}{\rm ln}[
M_G({\cal M}^{-1})_{11}] \nonumber \\
&+&{{\alpha}\over{60\pi}}\left\{21{\rm ln}(1+x^2) -{201\over 17}{\rm ln}
(4+x^2) -{156 \over 17}{\rm ln}x^2\right\}
\end{eqnarray}

\noindent
The uncertainty in sin$^2\theta_W$ is $\pm 0.0033 \pm 0.0014 \pm 0.0006$
[2], where the first number corresponds to the experimental errors in
$\alpha (M_Z)$ and $\alpha_3(M_Z)$, the second one to SUSY particle
threshold and the last one to the top and higgs thresholds.

The terms in the first line of Eq. (24) is unambiguous and
adds up to $0.2415$.   Unless
there is some cancellation from the other terms, sin$^2\theta_W$ will be
incompatible with experiments.  The ln($2\beta_1^\prime$) term has an
extremely small coefficient and is negligible, the second term on
line 2 is, as noted earlier, tends to be
a positive contribution (see Eq. (8)).  The
last term in line 2 is nearly zero or positive.  As for the terms in the
curly brackets in the last line,
it is positive for $x < 0.5$, but it can be negative for
larger $x$, with its minimum being $-1.9 \times 10^{-4}$.  The
term with ln$(M_G({\cal M}^{-1})_{11})$ is probably positive (if the proton
is not to decay too fast) and is of order $10^{-3}$ at most.
That leaves us with the ln$\left(\delta^{(1)}_{eff}/\delta_{eff}^{(2)}\right)$
term to be the only term that can be significantly negative.  In its absence,
sin$^2\theta_W$ would come out too large.  But this term, depending on
the unknown ratio of $\delta$'s can bring sin$^2\theta_W$ to agreement
if the logarithm is about 3.  It should be emphasized that the two
$\delta$'s are $independent$ parameters of the model and can easily
differ from each other by some factor of order unity.

\section*{IV. Conclusions}

In this paper we have presented a very simple scheme for the gauge symmetry
breaking in the context of
SUSY-$SO(10)$ GUT.  The Higgs system employed consisting of a
single {\bf 45} along with a ${\bf \overline{16}}+{\bf 16}$ and a pair
of {\bf 10}, is the absolute
minimum required for symmetry breaking and a natural
doublet--triplet splitting without fine--tuning of parameters.
Such a spectrum has been shown to arise in the free fermionic formulation of
superstrings [11].
The mass
of the light Higgs doublet is protected by local symmetry
against higher dimensional
operators induced by Planck scale physics to sufficiently high order.
The model presented here is a simplification over
earlier attempts along these lines [8,9,14].

Since the symmetry breaking mechanism relies on non--renormalizable
operators (without such operators $SO(10)$ can only break down to
$SU(5)$), some of the Higgs(ino)s in the model turn out to have masses
below the GUT scale.  These pseudogoldstone multiplets affect
sin$^2\theta_W$ as well as the unification scale $M_U$.
(In Refs. [8,9], it was required that the spectrum below the GUT scale
should be the same as the MSSM spectrum,
so sin$^2\theta_W$ and $M_U$ predictions were
essentially the same as in minimal SUSY--$SU(5)$.  The price to be paid
was the necessity of having three adjoints.)  $M_U$ is found
to be about $2 \times 10^{17}~GeV$, which is closer to the string
compactification scale, while sin$^2\theta_W$ is somewhat on the large
end of the presently allowed range.  Threshold corrections to
sin$^2\theta_W$ turn out to be quite small, but they are large enough to make
the prediction consistent with experiments.  We have also shown that
these pseudogoldstone Higgs(ino)s do not pose any problem for
cosmology, as they decay sufficiently fast in the early universe.  They
also do not mediate proton decay at an unacceptable level.

The simplicity of the Higgs sector of the model
also means that all the couplings
will remain perturbative in the momentum range from $M_{GUT}$ to $M_{Pl}$.
Realistic fermion masses including small neutrino masses can arise naturally
within this scheme.  In fact, the model may prove to be a fertile ground for
implementing predictive schemes for quark and lepton masses.
This will be the subject of a future investigation.

\noindent
{\bf Acknowledgments:}

We wish to thank Rabi Mohapatra for discussions.  This work is supported
in part by a Grant from the Department of Energy.

\section*{References}

\begin{enumerate}

\item U. Amaldi, W. de Boer, and H. Furstenau, Phys. Lett. {\bf B260},
447 (1991); P. Langacker and M.X. Luo, Phys. Rev. {\bf D 44}, 817
(1991); J. Ellis, S. Kelley, and D.V. Nanopoulos, Phys. Lett.
{\bf B260}, 131 (1991).
\item P. Langacker and N. Polonsky, Phys. Rev. {\bf D 47}, 4028 (1993).
\item E. Witten, Phys. Lett. {\bf 105B}, 267 (1981); D.V. Nanopoulos and
K. Tamvakis, Phys. Lett. {\bf 113B}, 151 (1982); S. Dimopoulos and H.
Georgi, Phys. Lett. {\bf 117B}, 287 (1982); L. Ibanez and G. Ross, Phys.
Lett. {\bf 110B}, 215 (1982); A. Sen, Fermilab-Pub-84/67 (unpublished).
\item A. Buras, J. Ellis, M. Gaillard and D.V. Nanopoulos, Nucl. Phys.
{\bf B135}, 66 (1985); H. Georgi, Phys. Lett. {\bf 108B}, 283 (1982); A.
Masiero, D.V. Nanopoulos, K. Tamvakis and T. Yanagida, Phys. Lett. {\bf
115B}, 380 (1982); B. Grinstein, Nucl. Phys. {\bf B206}, 387 (1982).
\item S. Dimopoulos and F. Wilczek, Report No. NSF-ITP-82-07, August 1981
(unpublished); R.N. Cahn, I Hinchliffe and L. Hall, Phys. Lett.
{\bf 109B}, 426 (1982).
\item K. Inoue, A. Kakuto and T. Takano, Prog. Theor. Phys. {\bf 75},
664 (1984); A. Anselm and A. Johansen, Phys. Lett. {\bf B200}, 331
(1988); Z. Berezhiani and G. Dvali, Sov. Phys. Lebedeev Inst. Reports
{\bf 5}, 55 (1989); R. Barbieri, G. Dvali and M. Moretti, Phys. Lett. {\bf
B312}, 137 (1993); G. Dvali and Q. Shafi, Bartol Preprint BA-94-11.
\item See eg., Ch. IV of {\it Unity of Forces in the Universe}, A. Zee (
1982, World Scientific Pub. Co. Inc. Singapore), and ``Grand unified
theories'', S.M. Barr, in {\it Encyclopedia of Physics}, Second Edition,
ed. Rita G. Lerner and George L. Trigg (1991, VCH Pub. Inc, New York) p.445.
\item K.S. Babu and S.M. Barr, Phys. Rev. {\bf D 48}, 5354 (1993).
\item K.S. Babu and S.M. Barr, Phys. Rev. {\bf D 50}, 3529 (1994).
\item D. Lewellen, Nucl. Phys. {\bf B337}, 61 (1990); A. Font, L.
Ibanez, and F. Quevedo, Nucl. Phys. {\bf B345}, 389 (1990).
\item S. Chaudhuri, S. Chung and J. Lykken, FERMILAB-PUB-94/137-T
(1994).
\item M. Srednicki, Nucl. Phys. {\bf B202}, 327 (1982).
\item Z. Berezhiani and R. Rattazzi, Nucl. Phys. {\bf B 407}, 249 (1993);
S.M. Barr, Phys. Rev. Lett. {\bf 64}, 353 (1990); G. Anderson,
S. Dimopoulos, L. Hall, S. Raby and G.D. Starkman, Phys. Rev. {\bf D 49},
3660 (1994).
\item D. Lee and R.N. Mohapatra, Phys. Lett. {\bf B324}, 376 (1994) and
Maryland Preprint UMD-PP-94-166 (1994);
J. Hisano, H. Murayama and T. Yanagida, Phys. Rev. {\bf D 49}, 4966
(1994).
\end{enumerate}

\end{document}